# Periodically Poled Aluminum Scandium Nitride Bulk Acoustic Wave Resonators and Filters for Communications in the 6G Era

*Izhar[1], M. M. A. Fiagbenu[1], S. Yao[1], X. Du[1], P. Musavigharavi[1,2], Y. Deng[1], J. Leathersic[3], C. Moe[3], A. Kochhar[3], E. A. Stach[4], R. Vetury[3], and R. H. Olsson III[1,]*

[1]Department of Electrical and Systems Engineering, University of Pennsylvania, Philadelphia, PA 19104 USA

[2]Department of Materials Science and Engineering, University of Central Florida, Orlando, FL 32816 USA

[3]Akoustis Inc., Huntersville, NC 28078 USA

[4]Department of Materials Science and Engineering, University of Pennsylvania, Philadelphia, PA 19104 USA

**Corresponding authors:**
Prof. Roy H. Olsson III
Email: rolsson@seas.upenn.edu
Office: +1(215) 898-6424

**Abstract**

Bulk Acoustic Wave (BAW) filters find applications in radio frequency (RF) communication systems for Wi-Fi, 3G, 4G, and 5G networks. In the beyond-5G (potential 6G) era, high frequency bands (>8 GHz) are expected to require resonators with high quality factor ($Q$) and electromechanical coupling ($k_t^2$) to form filters with low insertion loss and high selectivity. However, both the $Q$ and $k_t^2$ of resonator devices formed in traditional uniform polarization piezoelectric films of aluminum nitride (AlN) and aluminum scandium nitride (AlScN) decrease when scaled beyond 8 GHz. In this work, we utilized 4-layer AlScN periodically poled piezoelectric films (P3F) to construct high frequency (~17-18 GHz) resonators and filters. The resonator performance is studied over a range of device geometries, with the best resonator achieving a $k_t^2$ of 11.8% and a $Q_p$ of 236.6 at the parallel resonance frequency ($f_p$) of 17.9 GHz. These resulting figures of merit are ($FoM_1 = k_t^2 Q_p$ and $FoM_2 = f_p FoM_1 \times 10^{-9}$) 27.9 and 500 respectively. These and the $k_t^2$ are significantly higher than previously reported An/AlScN based resonators operating at similar frequencies. Fabricated 3-element and 6-element filters formed from these resonators demonstrated low insertion losses (IL) of 1.86 dB and 3.25 dB, and -3 dB bandwidths (BW) of 680 MHz (fractional BW of 3.9%) and 590 MHz (fractional BW of 3.3%) at ~17.4 GHz center frequency. The 3-element and 6-element filters achieved excellent linearity with in-band input third order intercept point (IIP3) values of +36 dBm and +40 dBm, respectively, which are significantly higher than previously reported acoustic filters operating at similar frequencies.

## Introduction

Recent advancements in radio frequency (RF) and mobile communications demand high data rates and small footprints which require resonators and filters with broader bandwidths and higher frequencies.[1,2] Modern communication standards such as Wi-Fi, 3G, 4G, and 5G have pushed the frequencies beyond traditional bands (< 2.6 GHz) to levels as high as 7 GHz, facilitating the accommodation of broader bandwidths.[3,4] In the beyond-5G (potential 6G) era, high frequency bands (> 7 GHz) will be utilized, which are expected to require resonators with high quality factor ($Q$) and electromechanical coupling ($k_t^2$), once RF spectrum use in these bands increases.[5] Acoustic resonators, such as surface acoustic wave (SAW) and bulk acoustic wave (BAW) resonators have been used predominantly for realizing filters for modern RF communication.[3,6] Scaling SAW resonators to higher frequency requires very thin and delicate interdigitated transducer (IDT) electrode patterns, which suffer from low fabrication yields, poor power handling, and large ohmic losses. For higher frequencies, BAW resonators and filters are preferred due to high $k_t^2$, low insertion loss (IL) and good selectivity. Aluminum nitride (An), due to its low dielectric loss tangent, acoustic damping and compatibility with fabrication using standard complementary metal-oxide-semiconductor (CMOS) tool sets, has been utilized to implement BAW resonators in the literature.[7–10] Alloying an with scandium (Sc) to form aluminum scandium nitride (AlScN) has shown better piezoelectric properties than Alun. For instance, the piezoelectric coefficient, $d_{33}$, of AlScN surpasses that of pure AlN by up to five times, showcasing a substantial enhancement in performance.[11–13] The frequency of BAW resonators is strongly dependent on its thickness.[14] Scaling BAW resonators to higher frequencies, for beyond-5G applications (>8 GHz), requires a significant reduction in thickness, impacting impedance matching.[15] Though the area of the resonator can be reduced to maintain its input impedance, the area-to-perimeter ratio (*A/p*) of the resonator decreases significantly, which leads to degradation in both $k_t^2$ and *Q*.[14] Thus, there is a demand for BAW devices that can operate at elevated frequencies while maintaining equitably thick piezoelectric layers to ensure optimal performance. One strategy to preserve the same layer thickness while achieving a higher operational frequency for BAW devices involves the implementation of periodically poled piezoelectric film (P3F). Barrera et al.[16] developed acoustic resonators and filters utilizing P3F lithium niobate (LiNbO3) achieving a resonator *Q* of 40 and $k_t^2$ of 44% at 19.8 GHz frequency, leading to filter insertion loss (IL) of 2.38 dB and fractional BW of 18.2% at 16.5 GHz. Similarly, Kramer et. al.[17] constructed a P3F LiNbO3 acoustic resonator which obtained $Q_s$ of 55 and $k_t^2$ of 40%. However, suspended LiNbO3 filters have yet to be commercialized and the initial literature reports show a limited linearity, with an in-band, input referred intercept point (IIP3) of +8 dBm in K-band.[16] Utilizing a CMOS compatible 2-layer

P3F AlScN, a 13.4 GHz BAW was constructed by Mo et al.,[18] that attained a $Q_s$ of 151 and $k_t^2$ of 10.7%. An 18.4 GHz BAW resonator fabricated using P3F AlScN is reported by Vetury et al.[19] that achieved a $Q_p$ of 260 and $k_t^2$ of 7.6%. This was the first time that a P3F AlScN resonator has been constructed in a commercial manufacturing process making it suitable for mass production. Kochhar et al.[20] reported X-band P3F AlScN resonators and filters achieving a $Q_p$ of 789 and $k_t^2$ of 10% at 10.72 GHz frequency leading to filter insertion loss (IL) of 0.7 dB and fractional BW of 4.7% at 10.72 GHz frequency. Izhar et al.[21] constructed a 20 GHz BAW resonator using a 3-layer P3F AlScN that achieved $k_t^2$ of 8.23 % and $Q_p$ of 160.

In this work, we report BAW resonators with different sizes (A/p) for filter applications using an as grown and electrically poled 4-layer P3F AlScN film. The resonators achieved a maximum $Q_p$ of 236.6 at a parallel resonance frequency ($f_p$) of 17.9 GHz and a maximum $k_t^2$ of 11.8%, which resulted in higher figures of merits ($FoM_1 = 27.9$ and $FoM_2 = 500$) compared to state-of-the-art AlN and AlScN K and Ku band resonators. Filters using different topologies (3-element and 6-elements) were constructed from the P3F AlScN BAW resonators achieving low IL of 1.86 dB and 3.25 dB, and -3 dB bandwidths (BW) of 680 MHz (fractional BW of 3.9%) and 590 MHz (fractional BW of 3.3%) at ~17.4 GHz center frequency with good power handling and linearity (IIP3 of >34 dBm).

**Design and Fabrication**

The structure and working mechanism of the P3F AlScN BAW resonator is illustrated in Fig. 1. The resonator is comprised of a 4-layer AlScN P3F, as grown and electrically poled, sandwiched between top and bottom molybdenum (Mo) electrodes. Under electrical excitation, the device experiences contraction and expansion, due to different polarities of the AlScN, inducing tensile ($+\sigma_y$) and compressive ($-\sigma_y$) stresses in the fourth thickness extensional (TE4) mode. This process produces acoustic waves within the device which resonate at a frequency of approximately,

$$f_{P3F} = \frac{v}{\lambda} = \frac{4v}{2t} \tag{1}$$

where $\lambda$, $v$ and $t$ respectively denote the acoustic wavelength, acoustic velocity, and thickness of the resonator. The P3F AlScN enabled the resonator to operate at four times higher frequency compared to a similar unpoled resonator,

$$f_{PF} = \frac{v}{\lambda} = \frac{v}{2t} \tag{2}$$

operating in the fundamental thickness extension mode (TE1) as depicted in Fig. 1(b). The P3F BAW resonators and filters are constructed from AlScN because the electromechanical coupling[22]

$$k_t^2 = \frac{d_{33}^2 E_3}{\varepsilon} \tag{3}$$

for a BAW resonator (neglecting the metal electrodes) depends on piezoelectric coefficient ($d_{33}$), modulus of elasticity ($E_3$) and dielectric constant ($\varepsilon$). The $d_{33}$ of $Al_{0.72}Sc_{0.28}N$ (~ 20 pm/V) is much higher than other piezoelectric materials such as AlN (3.9 pm/V) and ZnO (5.9 pm/V), therefore it enables the BAW resonators to achieve high $k_t^2$.[23,24]

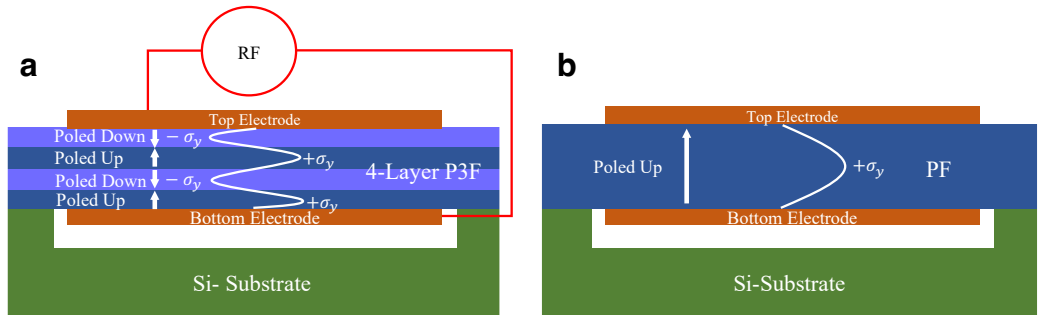

**Fig.1| Structure and working principle. a** P3F AlScN resonator operating in the 4[th] thickness extensional mode (TE4) with operating frequency 4 times higher than **b** similar unpoled resonator operating at the fundamental thickness extensional mode (TE1).

The resonators were fabricated by (1) the formation of as grown and periodically poled 4-layer AlScN P3F on a 6-inch silicon wafer, and (2) then subsequently transforming the P3F into BAW resonators through the utilization of the commercial XBAW$^{TM}$ process. The detailed process flow for the construction of the P3F AlScN 4-layer stack is described in Fig. 2(a). The P3F formation initiated with the deposition of 124 nm of $Al_{0.8}Sc_{0.2}N$ (layer1) in a Metal-polar (M-polar) orientation (poled down), utilizing the technique reported by Vetury et al.[19] and Moe et al.[25] (step I). Next, a Physical Vapor Deposition (PVD) co-sputtering process was employed to deposit 230 nm of $Al_{0.64}Sc_{0.36}N$ (layer 2) in a Nitrogen-polar (N-polar) orientation (step II). Following this, a 36 nm thick aluminum (Al) was sputtered in the same process chamber without a vacuum break to avoid oxide formation on the AlScN (step II). This was followed by patterning of the Al layer to form the bottom electrode for poling purposes. Another layer of 223 nm thick $Al_{0.64}Sc_{0.36}N$ (layer 3) in a N-polar orientation (poled up) was deposited via the PVD co-sputtering technique during the subsequent step (step IV). Following this, a 40 nm thick Al layer was sputtered using the same PVD system, and then patterned using wet etching chemistry to form the top electrode for electrical poling of $Al_{0.64}Sc_{0.36}N$ (layer 3). Using the Al electrodes, the top $Al_{0.64}Sc_{0.36}N$ (layer 3) was then electrically poled, transitioning from N-polar (poled up) to M-polar (poled down). Fig. 2(b) presents the current vs. applied voltage curve, demonstrating the switching of the top AlScN layer from N-polar orientation to M-polar orientation. The coercive voltage of approximately 150 V was identified during ferroelectric switching by the abrupt increase in current upon ferroelectric switching of the layer. The poling was achieved at the filter level where 3 and 6 BAW resonators were poled in a single poling step. The top Al was removed in the subsequent step (V). Finally, a 115 nm thick $Al_{0.64}Sc_{0.36}N$ (layer 4) was deposited via co-sputtering to realize the 4-layer AlScN P3F. The surface of the manufactured AlScN P3F underwent analysis using an atomic force microscope (AFM), revealing a surface roughness of < 2 nm.

The 6-inch wafers with the 4-layer AlScN P3F were sent to Akoustis Inc., where a commercial XBAW$^{TM}$ process was used to construct BAW resonators and filters from P3F. The cross-sectional depiction of a typical BAW resonator, accomplished through the XBAW$^{TM}$ process, is illustrated in Fig. 3. Detailed information on the XBAW$^{TM}$ process can be found in the pertinent literature.[19,26]

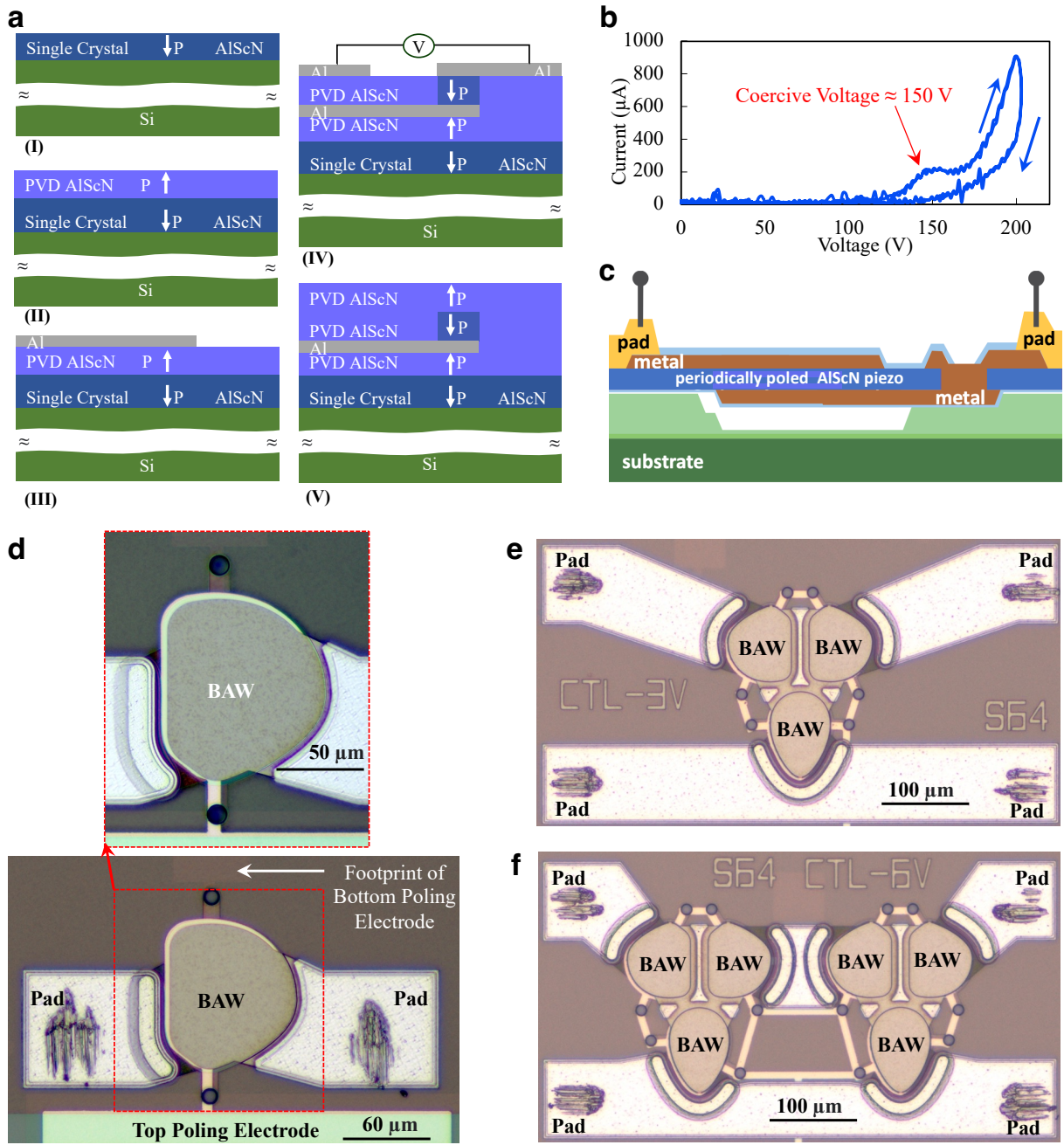

**Fig. 2 | Resonator and filter fabrication and poling waveform. a** The fabrication process for P3F AlScN involves: (I) deposition of an $Al_{0.8}Sc_{0.2}N$ (layer 1) in a M-polar orientation, (II) PVD deposition of $Al_{0.64}Sc_{0.36}N$ (layer 2) in a N-polar orientation, (III) sputtering of Al, serving as the bottom electrode for poling, (IV) PVD deposition of $Al_{0.64}Sc_{0.36}N$ (layer 3), Al top electrode for poling, and electrical poling of deposited $Al_{0.64}Sc_{0.36}N$ (layer 3), and finally, (V) stripping off the top Al poling electrode and PVD deposition of $Al_{0.64}Sc_{0.36}N$ (layer 4), **b** the current vs. voltage response recorded during the electrical poling of the AlScN (layer 3) in step IV, showing the transition of AlScN (layer 3) from N-polar to M-polar. This waveform underlines the critical switching coercive voltage point, which registers at approximately 150 V, **c** cross-sectional view of the BAW resonator realized in XBAW process, **d** optical micrograph of a BAW resonator showing the device pads and poling electrodes, and optical micro image of the fabricated **e** 3-element and **f** 6-element filters.

## Results and Discussion

A vector network analyzer, VNA (Keysight Technologies E8361A PNA) was used to characterize the P3F resonators and filters. A Signal-Ground (SG) probe (ACP-40 SG 200, FormFactor Inc., USA) was used to connect the resonator to the VNA. One port measurements were taken with an input power of -10 dBm, a port impedance of 50 Ω, and a frequency range of 3 to 25 GHz. Calibration to the probe tips was accomplished using the short-open-load (SOL) technique.

The data extracted using the VNA contains the on-wafer interconnects to the GS probe in addition to the intrinsic resonator response. De-embedding was performed to obtain the response of the intrinsic resonator from the measured reflection factor ($\Gamma_M$) using equivalent open ($\Gamma_{M,open}$) and short ($\Gamma_{M,short}$) reflection factors produced on the same Si-wafer.[27]

The admittance responses of the resonators after de-embedding are shown in Fig. 3a. It is observed from the figure that the devices exhibit distinct vibration modes corresponding to the first (TE1), second (TE2), third (TE3) and fourth TE (TE4) modes at approximately 3.5 GHz, 8 GHz, 12 GHz and 18 GHz frequencies, respectively. The figure illustrates that the devices show a dominant response at TE4 (~17 GHz). This observation also confirms the successful poling of the devices, as depicted in Fig. 1, where the P3F devices were specifically engineered to operate optimally in the TE4 mode. To further clarify these findings, simulations were conducted using COMSOL Multiphysics to analyze the vibration modes of the devices. As shown in Fig. 3b, the stress distributions of the devices indicate their operation in the TE4 mode.

Furthermore, shunt resonators with slightly thicker metal electrodes were exploited towards the implementation of ladder filters, with measured responses shown in Fig. 3d-e. These resonators also operated in the TE4 mode, which is clear from the mode shapes shown in Fig. 3f. However, due to the thick top metal electrodes, these resonators have slightly lower resonant frequencies ($f_p$ ~ 17.25 GHz) than the series resonators ($f_p$~18 GHz), which are utilized to realize broadband filters.

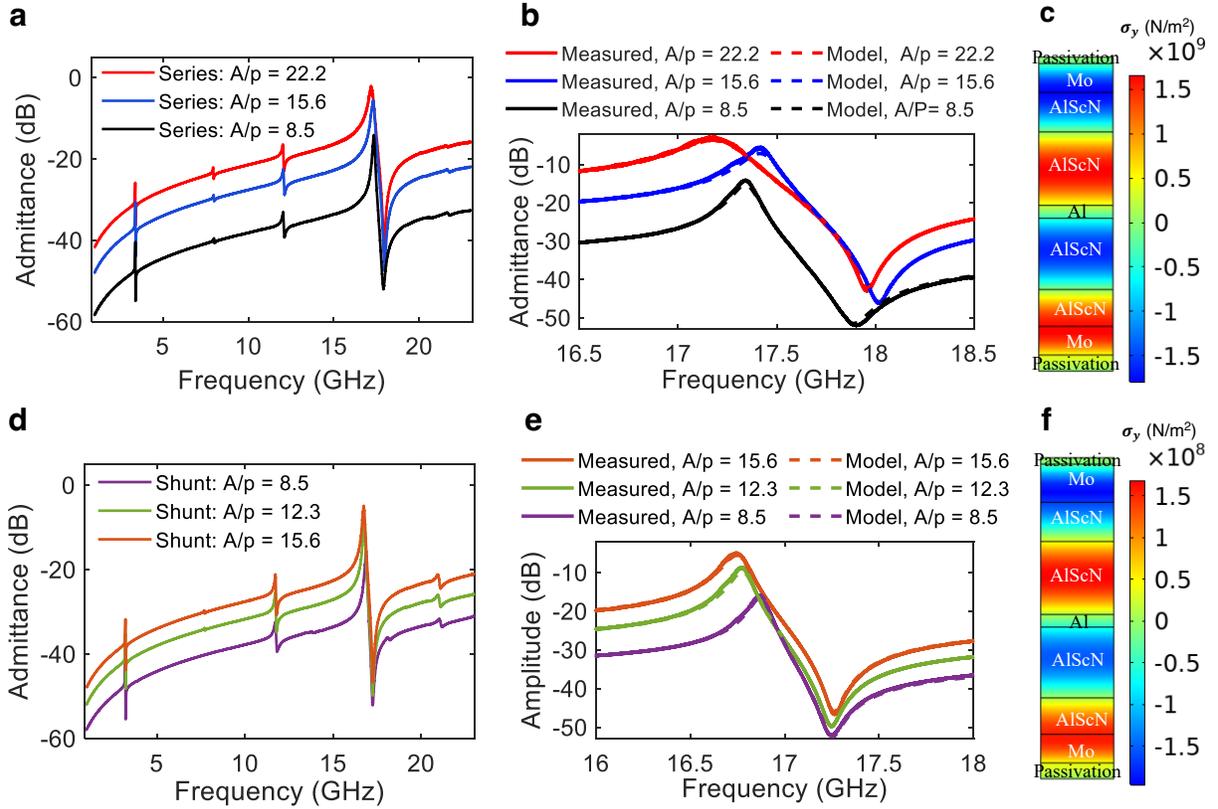

**Fig. 3 | Measured results and simulated mode shapes of resonators.** Admittance responses of **a-b** series and **d-e** shunt resonators showing their operation at the fourth thickness extensional mode (TE4). The stress distribution of the **c** series and **f** shunt resonators confirms the TE4 mode of vibration for the devices.

A modified Butterworth Vandyke (mBVD) model, as presented in Table I, consisting of a series resistance ($R_s$), a motional resistance ($R_m$), capacitance ($C_m$) and inductance ($L_m$), and a shunt capacitance ($C_0$) and resistance ($R_o$), is used to fit the measurements results of both series and shunt resonators. The measurement results in terms of admittance response of both series and shunt resonators are in close agreement with the mBVD model as shown in Fig. 3b and Fig. 3e, respectively. All the devices showed very clean responses in their region of operation and no unwanted spurs were observed. Devices with higher A/p ratio exhibited higher admittance which is due to their higher values of shunt capacitance compared to devices with smaller A/p ratio.

The Bode Q[28,29]

$$Q(\omega) = \omega \frac{d\varphi}{d\omega} \frac{|S_{11}|}{1-|S_{11}|^2} \tag{4}$$

of the devices were calculated in terms of S-parameter ($S_{11}$), phase ($\varphi$) and frequency ($\omega$). The Bode Q of both the measurement data and mBVD model fit very closely as provided in supplementary data.

Table 1 | The mBVD model fitted parameters and values to experimental results for the series and shunt resonators.

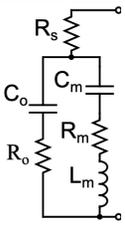

| mBVD Model | Resonator | A/p ratio | $R_s$ (Ω) | $C_o$ (pF) | $R_m$ (Ω) | $L_m$ (nH) | $C_m$ (fF) | $R_o$ (Ω) |
|---|---|---|---|---|---|---|---|---|
| | | 22.2 | 1.142 | 1.2552 | 0.385 | 0.755 | 113.407 | 0.015 |
| | Series | 15.5 | 1.195 | 0.6042 | 1.28 | 2.013 | 41.436 | 0.022 |
| | | 8.5 | 1.132 | 0.1761 | 4.75 | 7.656 | 10.995 | 2.157 |
| | | 15.5 | 0.813 | 0.6137 | 1.1 | 2.384 | 37.863 | 0.052 |
| | Shunt | 12.3 | 0.971 | 0.3576 | 2.121 | 4.476 | 20.097 | 0.097 |
| | | 8.5 | 0.905 | 0.184 | 5.89 | 10.839 | 8.205 | 0.217 |

The performance parameters for all the devices in terms of parallel quality factor ($Q_p$), series quality factor ($Q_s$), and electromechanical coupling ($k_t^2$)

$$k_t^2 = \frac{\pi^2}{8} \left[ \frac{f_p^2 - f_s^2}{f_s^2} \right] \qquad (5)$$

are summarized in Fig. 4a, Fig. 4b and Fig. 4c, respectively. It is clear from Fig. 4a that the $Q_p$ increases with an increase in the A/p ratio for the devices. This is because with an increase in the A/p ratio, the acoustic loss to the anchor boundaries also decreases which helps improve the $Q_p$.[30] A maximum $Q_p$ of 236.6 is achieved by the series resonator with an A/p ratio of 22.2. While the $Q_p$ increases, the $Q_s$ of both series and shunt resonators decreases with increasing A/p ratio as shown in Fig. 4b. This is because, while $Q_p$ is dominated by the acoustic loss, the $Q_s$ is dominated by the ohmic loss. Since the resonators with higher A/p ratio have smaller motional resistances, therefore, they are impacted more by the series resistance hence lowering their $Q_s$.[30] The $Q$ of shunt resonators is slightly higher than the series resonators for the same A/p ratio. This is because of the mass loading effect of the thick Mo top electrodes which helped the shunt resonators achieve higher $L_m$ compared to the series resonators with thin electrodes and the reduced $R_s$ from the thicker metal (see Table 1), thus improving their $Q_p$ and $Q_s$. The $k_t^2$ of the resonators, as shown in Fig. 4c, increases with an increase in the A/p ratio. This is because the center part of the resonator (proportional to A) is free to move while the periphery of the resonator (proportional to p) is attached to the substrate and does not vibrate. Therefore with an increase in A/p ratio more electrical energy is converted into acoustic energy within the resonator thus improving $k_t^2$.[30] A maximum $k_t^2$ of 11.8% is achieved by the series resonator with an A/p ratio of 22.2, which resulted in a figure of merit ($FoM_1=k_t^2 \cdot Q_p$) of 27.9 for the same resonator.

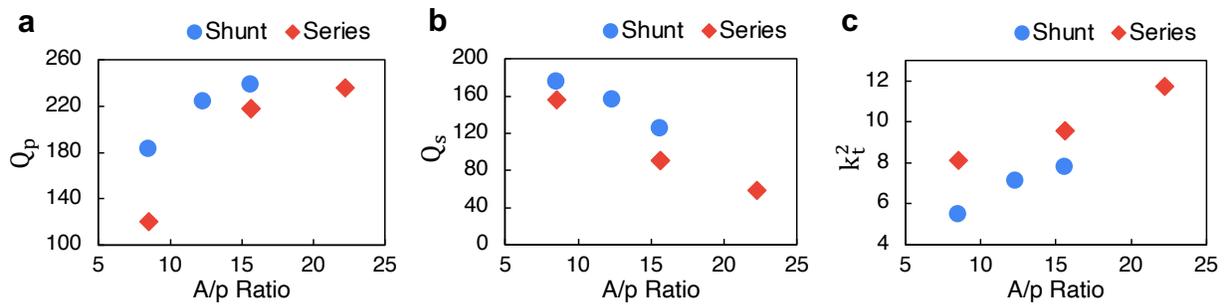

**Fig. 4 | Performance parameters of the resonators as a function of their dimension. a** $Q_p$ of both series and shunt resonators increases whereas **b** $Q_s$ decreases with an increase in the A/p of the device. **c** $k_t^2$ of both series and shunt resonators increase with increasing device A/p.

High frequency bandpass filters using different topologies (such as 3-element and 6-element) were constructed from the P3F AlScN BAW resonators as shown in Fig. 5a and Fig. 5b, respectively. The 3 and 6 element filters achieved a low IL of 1.86 dB and 3.25 dB, and -3 dB bandwidths (BW) of 680 MHz (fractional BW of 3.9%) and 590 MHz (fractional BW of 3.3%) at ~17.4 GHz center frequency. Furthermore, the 3 and 6 element filters achieved out-of-band rejection of approximately 7 dB and 16.6 dB, respectively. The power handling and linearity of the filters were also analyzed, revealing in-band IIP3 values of +36 dBm and +40 dBm for the 3 and 6 element filters, respectively, as depicted in Fig. 5e and Fig. 5f. The measured values are limited by the experimental setup as an IIP3 of similar value is obtained for a standard thru structure on a calibration substrate, as shown in Fig. 5g.

The performance of the P3F resonators and filters reported in this work are compared with the existing state-of-the-art in Table 2. The $k_t^2$ and FoM obtained of 11.8% and 27.9 are much higher than that previously reported for AlN/AlScN based resonators operating in K and Ku bands, owing to the increased number of P3F layers and the use of 36% Sc alloyed AlScN. Further, the high frequency filters (~18 GHz) showed much higher linearity and higher out-of-band rejection when compared to the existing state-of-the-art filters. While the FoM of LiNbO$_3$ resonators are higher due to the larger $k_t^2$, the higher Q of AlScN P3F resonator technology makes it more appropriate for realizing low loss filters with narrower passbands < 8%. Also, LiNbO$_3$ resonators are not fully compatible with CMOS technology.

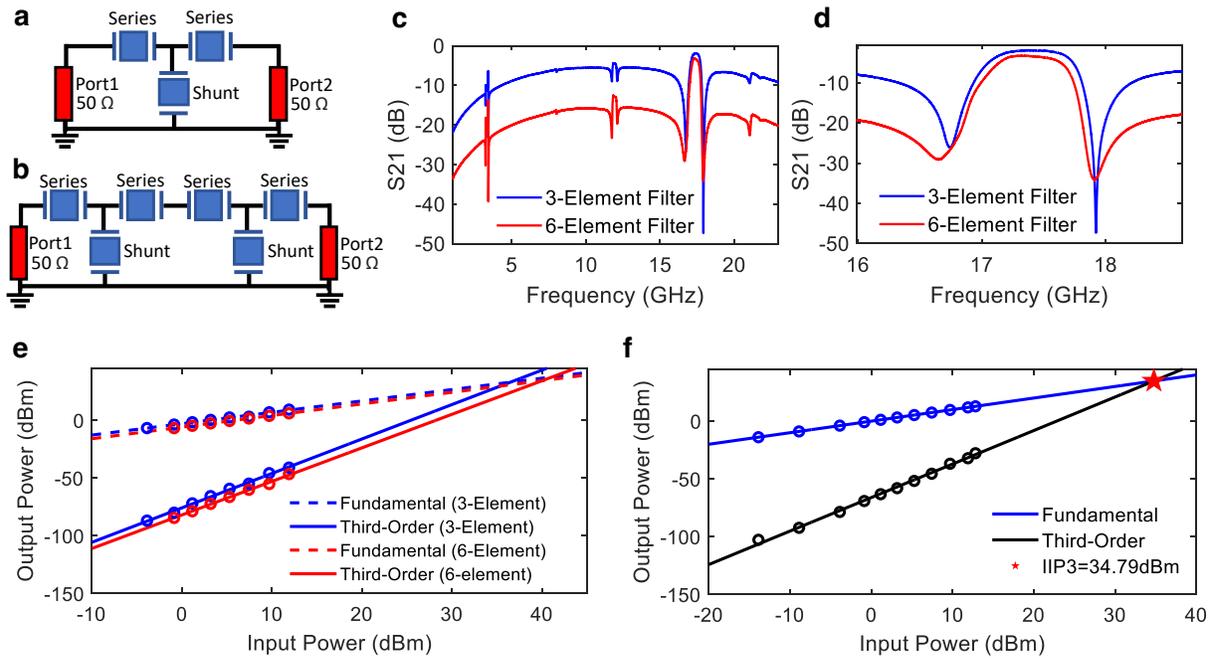

**Fig. 5 | Measured results of 3 and 6-element P3F filters**. Schematic of **a** three and **b** six elements filters. Broadband **c** and **d** narrow band response of the 3 and 6 element filters showing IL of 1.86 dB and 3.25 dB, and -3 dB bandwidths (BW) of 680 MHz (fractional BW of 3.9%) and 590 MHz, respectively. Input and output power of the filters showing **e** In-band IIP3 values of 36 and 40 dBm for the three and six element filters, **f** The response of a standard thru structure on a calibration substrate showing that our IIP3 measurement is limited by the experimental setup.

Table 1 | Comparison of the developed resonators and filters with the existing state of the art.

| Reference | P3F Material | Fully CMOS Compatible | Resnator Performance | | | | | Filter | | | |
|---|---|---|---|---|---|---|---|---|---|---|---|
| | | | Frequency (GHz) | $Q_s$ | $Q_p$ | $k_t^2$ (%) | $FoM_1$ | IL (dB) | BW (%) | Rejection (dB) | Linearity (dBm) |
| Lu et al.[31] (2020) | LiNbO$_3$ | Non-CMOS | 19.5 | NR | 445 | 1.26 | 5.6 | 7.86 | 0.21 | - | - |
| Kramer et. al.[17] (2022) | LiNbO$_3$ | Non-CMOS | 19.2 | 55 | - | 40 | 22 | - | - | - | - |
| Barrera et al.[16] (2023) | LiNbO$_3$ | Non-CMOS | 19.8 | 40 | - | 44 | 17.6 | 2.38 | 18.2 | 13 | 8 |
| Barrera et al.[32] (2024) | LiNbO$_3$ | Non-CMOS | 38.2 | 13 | - | 30 | 38.7 | 5.63 | 17.6 | 10.2 | - |
| Mo et al. [33] (2022) | AlScN | CMOS | 13.9 | 151 | NR | 10.7 | 16.2 | - | - | - | - |
| Kochhar et al.[20] (2023) | AlScN | CMOS | 10.72 | 347 | 789 | 10 | 79 | 0.7 | 4.7 | 6 | - |
| Izhar et el.[21] (2023) | AlScN | CMOS | 20 | 92 | 160 | 8.23 | 13.2 | - | - | - | - |
| Vetury et al.[19] (2023) | AlScN | CMOS | 18.4 | 180 | 260 | 7.6 | 20 | - | - | - | - |
| Schaffer et al.[34] (2023) | AlN | CMOS | 55.7 | 93 | 89 | 2.2 | 2.1 | - | - | - | - |
| Cho et al.[35] (2023) | AlScN | CMOS | 21 | 62 | 32 | 6.4 | 2.1 | - | - | - | - |
| | AlScN | CMOS | 55.4 | 30 | 27 | 3.8 | 1.1 | - | - | - | - |
| **This Work** | **AlScN** | **CMOS** | **17.4** | **58.4** | **236.6** | **11.8** | **27.9** | **3.25** | **3.4** | **16.6** | **>40 dBm** |

## Conclusion

In this study, we employed P3F AlScN to fabricate resonators and filters operating at high frequencies (~18 GHz). The resonator with the highest A/p ratio achieved a $Q_p$ value of 236.6 at a $f_p$ of 17.9 GHz and a $k_t^2$ value of 11.8%. This led to figures of merit ($FoM_1 = k_t^2 Q_p$ and $FoM_2 = f_p FoM_1 \times 10^{-9}$) of 27.9 and 500 for the resonator, respectively. These figures of merit surpass those of existing state-of-the-art resonators realized in AlN and AlScN materials operating at similar or higher frequencies. The 3-element and 6-element filters exhibited low IL of 1.86 dB and 3.25 dB, along with -3 dB bandwidths (BW) of 680 MHz (3.9% fractional BW) and 590 MHz (3.3% fractional BW) at a center frequency of ~17.4 GHz. Both filters demonstrated excellent linearity, with in-band IIP3 values of >40 dBm. It is concluded from the experimental results that P3F AlScN resonator and filter technology holds promise for advancing RF communications in the beyond-5G era.

## Methods and Materials

### Microfabrication of periodically poled piezoelectric film (P3F)

The periodically poled piezoelectric film (P3F) was produced on 6 inch Silicon wafers at the University of Pennsylvania. The wafer was first sent to the AKTS foundry for printing alignment features. This was done to align the in-house processing steps to those performed in the AKTS foundry. The first layer deposition of 124 nm of $Al_{0.8}Sc_{0.2}N$ (layer1) in a M-polar orientation (poled down) was also performed in the AKTS foundry. The wafers received from the foundry were processed in a PVD system to deposit subsequent layers of AlScN. In the PVD system, the targets underwent conditioning to ensure the deposition of high-quality AlScN. Utilizing separate 4-inch diameter Al and Sc targets, along with nitrogen gas, facilitated the deposition process. Prior to depositing each AlScN layer, the process chamber underwent pumping to attain a base pressure of approximately $9x10^{-8}$ mbar, and the substrate temperature was elevated to 350°C. To attain the desired composition of 36% Sc in the deposited film, the respective cathode powers for the Sc and Al targets were set at 655 W and 900 W during deposition, with a nitrogen flow rate of 20 sccm.

For the patterning of top and bottom Al poling electrodes, Microposit®s1813® photoresist was used. The photoresist was spin coated at 3000 rpm and baked at 115°C for 1 minute before exposing in a contact lithography tool. The resist was developed using microposit MF CD-26 developer at room temperature. To protect the thin film Al from the possible attack of photoresist developer (MF CD-26 developer), a layer of polymer (ZEP 520A) was spin coated at 2500 rpm and baked at 120°C before coating and developing the photoresist. After the photoresist development, the polymer (ZEP 520A) was etched using $O_2$ plasma to expose the thin film Al. Both the masking layers (polymer and photoresist) after patterning of Al were cleaned using microposit remover 1165 at 65°C.

### De-embedding Technique Used to Obtained the Resonators Intrinsic Response from the Measured Data

It should be noted that the data extracted using the VNA contain the interconnects to the GS probe in addition to the intrinsic resonator response. De-embedding was performed to obtain the reflection factor of the intrinsic resonator

$$\Gamma_{DUT} = \frac{A - \Gamma_M}{A^2 - A.\Gamma_M - B} \tag{6}$$

from the measured reflection factor ($\Gamma_M$) using the parameters

$$A = \frac{\Gamma_{M,open} + \Gamma_{M,short}}{2 + \Gamma_{M,open} - \Gamma_{M,short}} \tag{7}$$

and

$$B = (\Gamma_{M,open} - A)(1 - A) \tag{8}$$

that were extracted using equivalent open ($\Gamma_{M,open}$) and short ($\Gamma_{M,short}$) reflection factors produced on the same Si-wafer.[27]

## Acknowledgements

The authors would like to thank Dr. Todd Bauer and Dr. Benjamin Griffin for their support of this work under the Defense Advanced Research Project Agency (DARPA) COmpact Front-end Filters at the ElEment-level (COFFEE) Program. The BAW resonators were partly fabricated at the Singh Center for Nanotechnology, which is funded under the NSF National Nanotechnology Coordinated Infrastructure Program (NNCI-1542153). The views, opinions and/or findings expressed are those of the author and should not be interpreted as representing the official views or policies of the Department of Defense or the U.S. Government.